\documentclass[preprint,bm,cite]{epl}

\usepackage{bm}
\usepackage{graphicx}
\usepackage{amssymb,amsmath}

\shrinkacknowledgements{1}
\shrinkfloats{1}

\newcommand{\smneq}{\! \neq \!}

\newcommand{\ve}{\varepsilon}

\newcommand{\Ef}{E_{\mathrm{F}}}

\newcommand{\be}{\begin{equation}}
\newcommand{\ee}{\end{equation}}
\newcommand{\bea}{\begin{eqnarray}}
\newcommand{\eea}{\end{eqnarray}}

\newcommand{\ci}{\mathrm{i}}

\title{Spin Accumulation and Equilibrium Currents at the Edge of 2DEGs with 
Spin-Orbit Coupling}
\shorttitle{Spin Accumulation and Equilibrium Currents}
\author{Gonzalo Usaj \and C. A. Balseiro}
\institute{Instituto Balseiro and Centro At\'{o}mico Bariloche, Comisi\'{o}n 
Nacional de Energ\'\i a At\'{o}mica, (8400) San Carlos de Bariloche, Argentina.}


\begin{document}
\maketitle

\begin{abstract}
We show that in a two dimensional electron gas (2DEG) the interplay between the Rashba spin-orbit coupling 
and a potential barrier (\textit{eg.}, the sample edge) generates interesting spin effects.
In the presence of an external charge current,  a \textit{spin accumulation} is built up near the barrier while in equilibrium there is an inhomogeneous spin current which is localized at the barrier and flows parallel to it. 
When an in-plane magnetic field perpendicular to the barrier is applied, the system also develops an inhomogeneous {\it charge 
current} density. These effects originate purely from the structure of the eigenstates near the boundary.
\end{abstract}

Spintronics has recently emerged as an active field of research due 
to its potential impact on both the design of future electronic 
devices and quantum computing \cite{Spintronicsbook}. The ultimate goal of spintronics is the manipulation 
and coherent control of the electronic spin degrees of freedom. This requires the 
ability to generate, inject and control spin-polarized currents in electronic 
devices---an example of that is the spin-transistor proposal by Datta and Das 
\cite{DattaD90}. In recent years, the spin-orbit coupling has been recognized as 
an efficient tool to manipulate the electron's spin leading to a substantial amount of 
work devoted to study its effect on the transport properties 
of nanostructures and two dimensional electron gases (2DEG) 
\cite{MorozB99,GovernaleZ02,BellucciO03,MillerZMLGCG03,StredaS03,MishchenkoH03,SchliemannL03,EguesBL03,UsajB04_focusing,RokhinsonLGPW04,MishchenkoSH04}. 
An important step along this line was the recently reported observation of the predicted 
spin Hall effect (SHE) \cite{KatoMGA04,WunderlichKSJ05}: when a transport current 
flows through the system the opposite spins accumulate at the lateral edges of the 
sample. It has been argued that the SHE can be an extrinsic effect when it relies 
on impurity scattering \cite{DyakonovP71,Hirsch99} or an intrinsic effect when it 
originates from the spin-orbit coupling in a 2DEG \cite{SinovaCSJM04,MurakamiNZ03}.

In this Letter we show that in a ballistic 2DEG with Rashba spin-orbit coupling 
\cite{Rashba60,BychkovR84} the presence of a potential barrier in one 
direction leads to a \textit{current induced spin accumulation} at the sample edge. 
The effect is reminiscent of the intrinsic SHE although in our case it is based 
purely on geometric properties of the clean systems.
In addition, we show that there 
is an equilibrium spin current \cite{Rashba03}---ie. with no external fields applied---that 
flows pa\-ra\-llel to the barrier, and an \textit{inhomogeneous charge current density} when 
an in-plane magnetic field is applied. 
These intrinsic effects are relevant for a better understanding of spin-sensitive transport experiments
and numerical results in small systems \cite{ShengST05,NikolicSZS04,NikolicZS04,HankiewiczMJS04}. 

\begin{figure}[t]
\onefigure[width=.8\textwidth,clip]{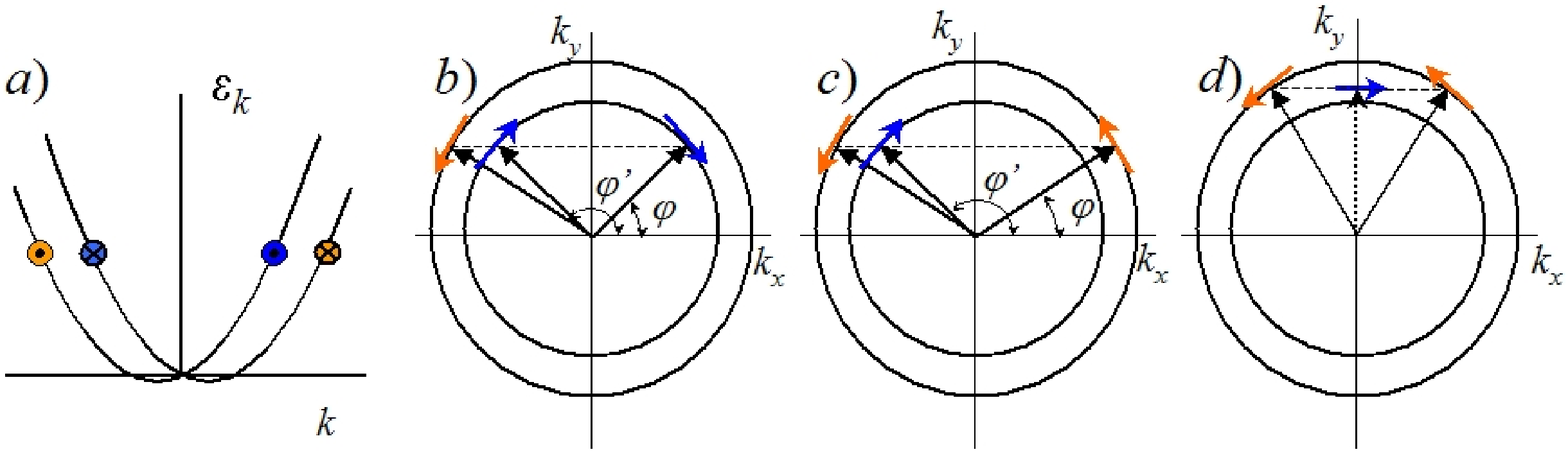}
\caption{Schematics of the band structure.
The dots show the spin orientation in the two bands. The others panels show
all possible configurations (for $\varepsilon >0$) of the incident plane
wave and the corresponding spin orientation together with the two Fermi
surfaces: b) incident plane wave in the `$+$' band; c) incident plane wave
in the `$-$' band with $|k_{y}|<k_{+}$, and d) with $|k_{y}|>k_{+}$. In
the latter case, the reflected component on the `$+$' band is an
evanescent mode whose spin points out of the plane.}
\label{scheme}
\end{figure}

We consider a 2DEG with Rashba spin-orbit coupling. 
The Hamiltonian is then given by 
\begin{equation}
H=\frac{p_{x}^{2}+p_{y}^{2}}{2m^{*}} +V(\bm{ r})+ \frac{\alpha }{\hbar }%
\left( p_{y}\sigma _{x} - p_{x}\sigma _{y}\right) \,,  \label{RashbaH}
\end{equation}
where $V(\bm{ r})$ is the confining potential in the plane of the 2DEG, 
$m^{*}$ is the effective mass and $\alpha $ is the
Rashba coupling constant \cite{NittaATE97,MillerZMLGCG03}.
For a system with translational invariance ($V(\bm{ r})=0$), the
eigenfunctions of Hamiltonian (\ref{RashbaH}) are 
given by 

\begin{equation}
\Psi _{\pm }(\bm{ r})=\frac{1}{\sqrt{2A}}{\rm e}^{\ci\,\bm{k\cdot r}%
}\left( {%
{\pm {\rm e}^{-\ci\phi /2} \atop {\rm e}^{\ci\phi /2}}%
}\right) \,,  
\label{wavefunction}
\end{equation}
with ${\rm e}^{\ci\phi }=(k_{y} - \ci k_{x})/k$ and $A$ the
system's area. Notice that if we write $\bm{k}=k(\cos \varphi
,\sin \varphi )$, then $\phi =\varphi  - \pi /2$. Consequently, these
modes have the spin in the plane of the 2DEG pointing in a direction
perpendicular to the wavevector $\bm{k}$ as schematically shown in Fig. 
\ref{scheme}. The spin degeneracy is lifted by the spin-orbit coupling
leading to two non-degenerated bands with the following dispersion relation: 
$\varepsilon _{\pm }(\bm{k })=\hbar ^{2}k^{2}/2m^{*} \pm  \alpha k$.

Let us now consider a system with a sharp edge in the $x$-direction defined
by the potential $V(\bm{ r}) \equiv  V_{0}\Theta (x)$ with $\Theta (x)$
the step-function. For $V_{0} \gg  E_{\mathrm{F}}$, where $E_{\mathrm{F}}$ is
the Fermi energy, the boundary condition imposes the eigenfunc\-tions with $\ve_{\pm} \leq \Ef$ to
be zero for $x \geq  0$ (the exponential tail inside the barrier is
negligible). For $x \leq  0$ they can be written as superposition of
incident and reflected waves and for positive energy $\varepsilon$ we have:
\begin{equation}
\Psi _{\bm{ k},s}(\bm{ r})=\frac{{\rm e}^{\ci\,k_{y}y}}{C^{\frac{1}{2%
}}}   \left\{  {\rm e}^{\ci\,k_{x}x}\left( {%
{s{\rm e}^{ - \ci\phi /2} \atop {\rm e}^{\ci\phi /2}}%
}\right) +R\,{\rm e}^{-\ci\,k_{x}x}\left( {%
{s{\rm e}^{\ci\phi /2} \atop {\rm e}^{-\ci\phi /2}}%
}\right) \right. \left. +R^{\prime }\,{\rm e}^{-\ci\,k_{x}^{\prime }x}\left( {%
{-s{\rm e}^{-\ci\phi ^{\prime }/2} \atop {\rm e}^{\ci\phi ^{\prime }/2}}%
}\right) \right\} \,,  
\label{band}
\end{equation}
where $s=\pm 1$ is the band index of the incident plane wave with
wavevector $(k_{x},k_{y})$, 
$C$
is a normalization constant, $(-k_{x},k_{y})$ and $(-k_{x}^{\prime
},k_{y})=k_{ - s}(\cos \varphi ^{\prime },\sin \varphi ^{\prime })$
are the wavevectors of the reflected plane waves, $k_{s}=[(2\hbar
^{2}\varepsilon /m^{*} + \alpha ^{2})^{\frac{1}{2}} - s\alpha
]m^{*}/\hbar ^{2}$ and 
\begin{equation}
R=-\frac{\cos (\left[ \varphi  - \varphi ^{\prime }\right] /2)}{\sin
(\left[ \varphi  + \varphi ^{\prime }\right] /2)},\quad R^{\prime }=%
\frac{\ci\cos \varphi }{\sin (\left[ \varphi  + \varphi ^{\prime
}\right] /2)}\,.  \nonumber  
\label{coeficients}
\end{equation}
For $\varphi\smneq0$, the reflected wave has a non-zero amplitude in both bands as illustrated in
Fig.\ref{scheme} (this is a simple way to create polarized beams out of an 
unpolarized one \cite{KhodasSF04,ChenHPDG05}). When the incident plane wave is on the 
$s=-1$ band, the above solution remains valid only if $|k_{y}| \leq k_{+}$
For $|k_{y}| > k_{+}$ there are no propagating modes in the ``$+$'' band and evanescent 
modes localized at the boundary appear (Fig.1d). In that case $k_{x}^{\prime }$ is 
replaced by an imaginary wavevector and the corresponding spinor acquires an explicit 
projection on the $z$ axis: 
\be
\Psi_{\bm{k},-}(\bm{r})=\frac{{\rm e}^{\ci\,k_{y}y}}{C'^{\frac{1}{2}}}
 \left\{ {\rm e}^{\ci\,k_{x}x} \left( {{ -{\rm e}^{ - \ci\phi/2} \atop {\rm e}^{\ci\phi/2}}
}\right)+{\rm e}^{\ci\,\Theta }\,{\rm e}^{-\ci\,k_{x}x} \left( {%
{ -{\rm e}^{\ci\phi/2} \atop {\rm e}^{-\ci\phi/2}}%
}\right) \right.
\left. + R_{ev}\,{\rm e}^{Kx}\left( {
{a \atop b}
}\right) \right\} \,,
\label{evanescent}
\ee
with $K=(k_y^2- k_{+}^{2})^{\frac{1}{2}}$ and $a^2/b^2=(k_y+K)/(k_y-K)$.
Since the incident and the reflected plane waves have different spin
projections, their interference leads to a non-zero spin density on the $z
$-axis, $\langle {\hat{\sigma}}_{z}\rangle _{\bm{ k},s}=\Psi _{\bm{ k}%
,s}^{\dagger }(\bm{ r}){\hat{\sigma}}_{z}\Psi _{\bm{ k},s}(\bm{ r})$ with
some spin accumulation at the sample edge ($x \lesssim  0$). The sign of the spin
accumulation depends on the sign of the conserved component $k_{y}$ of the
wavevector. Consequently, in the ground state where states with positive and
negative values of $k_{y}$ are occupied, the contribution to the local
magnetization of each band is zero. 
However, if there were a
preferential motion of the carriers in the $y$-direction---imposed by an
external current, for instance---there would be a net magnetization at the
surface. To illustrate this effect, we consider an ideal ballistic system
where the voltage drops at the contacts and the electric field is zero at
the interior of the sample. Then, in the energy interval [%
$E_{\mathrm{F}}+ eV/2,E_{\mathrm{F}}- eV/2$], only states with $k_{y} \geq 0$
are occupied within the sample \cite{Dattabook} and we obtain (for $E_{\mathrm{F}}>eV/2$): 
\be
\langle {\hat{\sigma}}_{z}(x)\rangle =
\sum_{\bm{k},s;k_{y}>0}\langle {\hat{\sigma}}_{z}\rangle_{\bm{k},s}\, 
F_{\bm{k},s}(E_{\mathrm{F}},eV)  
\label{spinacc}
\ee
with $F_{\bm{k},s}(E_{\mathrm{F}},eV)=\Theta
(E_{\mathrm{F}}+ eV/2-\varepsilon_{s}(\bm{k}))-\Theta (E_{\mathrm{F}}- 
eV/2-\varepsilon_{s}(\bm{ k}))$ \cite{note1}. For small $eV$ only states at $\Ef$ 
contribute to the sum and we may separate the contribution of the states with 
evanescent waves (Eq. (\ref{evanescent})) from the rest (Eq. (\ref{band})). 
For any $\Ef$, the former contribution can be scaled onto the $\Ef=0$ result 
(shown in Fig.2a) when plotted as a function of $(k^{2}_{-}-k^{2}_{+})^\frac{1}{2}x$. 
From this scaling law, it can be shown that the
integrated spin accumulation due to this contribution decreases with $%
E_{\mathrm{F}}$ as ($E_{\mathrm{F}}+\alpha ^{2}m^{*}/2 \hbar ^{2}$ )$^{-\frac{3}{4}}$%
. As we show below, states with reflected waves in the
two bands (Eq. (3)) tend to cancel the effect, leading to a faster decay of the 
spin accumulation as a function of $\Ef$. 
While the eigenfunctions presented above give a clear physical picture of the
edge effect, an analytical calculation is not simple since: a) the two modes of Eq. (\ref{band}) are not
orthogonal to each other; b) a proper counting of modes requires
to work with a system of finite width $L_{x}$. In what follows we
present results obtained by a numerical integration of the Schr\"{o}dinger
equation using finite differences. One advantage of the method is that it is
not restricted to a square well confining potential---our results hold in
the case of a sharp parabolic potential---and is equivalent to use a tight-
binding version of Hamiltonian (\ref{RashbaH}) \cite{Ando89}, 

\bea
H&=&\sum_{n\sigma }\varepsilon _{\sigma }c_{n\sigma }^{\dagger
}c_{n\sigma }-\sum_{<n,m>\sigma }t_{nm}c_{n\sigma }^{\dagger }c_{m\sigma
}+h.c.\nonumber \\  
&&-\lambda \sum_{n}\left\{ \ci\left(c_{n\uparrow }^{\dagger }c_{(n+%
\widehat{y})\downarrow }+c_{n\downarrow }^{\dagger }c_{(n+\widehat{y%
})\uparrow }\right) -\left( c_{n\uparrow }^{\dagger }c_{(n+\widehat{x})\downarrow
}-c_{n\downarrow }^{\dagger }c_{(n+\widehat{x})\uparrow }\right) \right\}
+h.c.
\eea
here $c_{n\sigma }^{\dagger }$ creates an electron at site $n$ with spin $%
\sigma _{z}=\sigma $ and energy $\varepsilon _{\sigma }=4t$, $%
t=\hbar ^{2}/2m^{*}a_{0}^{2}$, $a_{0}=5$nm is the lattice parameter, and $%
\lambda =\alpha /2a_{0}$. The summation is carried out on a square
lattice where the coordinate of site $n$ is $\bm{ r}_{n}=n_{x}\widehat{x}%
+n_{y}\widehat{y}$ with $\widehat{x}$ and $\widehat{y}$ the unit lattice
vectors.
 The hopping matrix
element $t_{nm}=t$ connects nearest neighbors. All
quantities presented below are obtained from the one particle propagators 
(for details see Ref.\cite{UsajB04_focusing}). We first present results for systems with
a large width $L_{x}$ and then we discuss the effect of the lateral confinement
in a quantum wire of reduced $L_{x}$.

\begin{figure}[t]
\onefigure[height=7.5cm,clip]{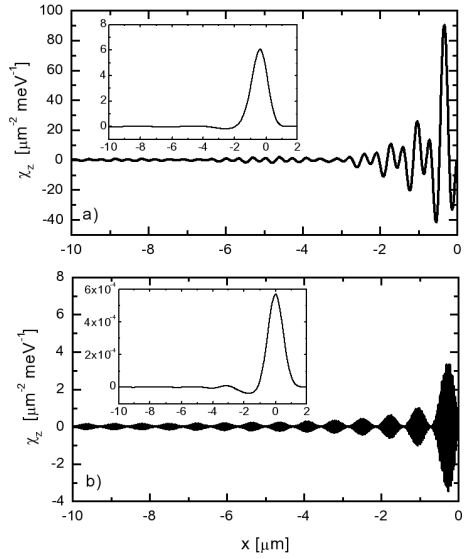} 
\caption{Current-induced spin accumulation, $\chi _{z}(x,\Ef)$, close to the potential barrier (edge) for $\Ef=0$ (a) and $\Ef=5$meV (b). 
The insets show the convolution with a Gaussian with an rms of $0.5\mu$m.}
\label{spincurrent}
\end{figure}

To analyze the spin accumulation we linearize Eq. (\ref{spinacc}) and
define $\chi _{z}(x)=\langle {\hat{\sigma}}_{z}(x)\rangle /eV$. 
The total spin accumulation for $\Ef=0$ is shown in Fig.2a. 
The inset shows the convolution with a Gaussian with an rms. of $0.5\,\mu$m.
For $\Ef > 0$ $(k_{+}\smneq0)$, the contribution from the states with $|k_{y}|<k_{+}$ partially cancels the effect as
shown in Fig. 2b. Thus, the maximum spin accumulation is obtained at $\Ef=0$ and as the carrier density increases it decreases and tends to
zero. $\chi _{z}(x)$ is strongly dependent on $\Ef$ so even for moderate values of $eV$, non-linear 
effects are observed and Eq. (\ref{spinacc}) should be used.

In this ballistic regime, the charge current also generates a local
magnetization $\langle {\hat{\sigma}}_{x}(x)\rangle $ in the $x$-direction (%
{\it i.e}. in the plane of the 2DEG and perpendicular to the sample edge) \cite{NikolicSZS04}
while by symmetry $\langle {\hat{\sigma}}_{y}(x)\rangle \equiv  0$ \cite{GovernaleZ02}---the non-zero value obtained for finite systems \cite{NikolicSZS04} is a consequence of the reflections at the sample-lead interfaces and depends strongly on the characteristics of such interface \cite{Reynosoetal}. 
\textit{It is important to emphasize that the spin accumulation at the edge is a purely geometric effect}. 
It is not directly connected to the one discussed in Ref. \cite{SinovaCSJM04}. The latter is due to an electric field induced
spin current, which is absent in the ballistic regime studied here.

Another consequence of the symmetry of $%
\langle {\hat{\sigma}_{z}}\rangle _{\bm{ k},s}$ with respect to $k_y$ is the presence of spin
currents along the edges of the sample. In equilibrium, we define the spin
current density as: 
\be
J_{y}^{\sigma _{z}}(x)=\frac{\hbar }{4} \sum_{\mathrm{occ}}\langle {\hat{v}}%
_{y}{\hat{\sigma}}_{z} + {\hat{\sigma}}_{z}{\hat{v}}_{y}\rangle _{k_y,\nu} 
=\frac{\hbar ^{2}}{2m^{*}} \sum_{\mathrm{occ}}k_{y}\langle {\hat{\sigma}}%
_{z}\rangle _{k_y,\nu}
\ee
here ${\hat{v}}_{y}=\ci/\hbar [{\hat{H}},{\hat{y}}]={\hat{p}_{y}}%
/m^{*} + \alpha {\hat{\sigma}}_{x}/\hbar $ is the $y$-component of the
velocity operator and the summation is restricted to all occupied states with quantum numbers $k_y$ and $\nu$.
 Here again the evanescent modes play a central role and $J_{y}^{\sigma
_{z}}(x)$ is {\it non-zero} in equilibrium.
Figs. 3a and 3b show the total spin current density $J_{y}^{\sigma
_{z}}(x)$ for different values of $\Ef$. The total spin current oscillates with a characteristic wavevector $
k_{F,+}+ k_{{F},-}$ and decays toward the bulk of the sample
with a characteristic length 
$(k_{F,-}- k_{{F},+})^{-1}=
\hbar^{2}/2m^{*}\alpha  \approx  100$nm that is energy
independent. This longitudinal spin current, integrated in half of the sample---
from $-L_{x}/2$ to $0$--- is non-zero: spin flows in opposite directions on
opposite sides of the sample. It is worth pointing out that this current does not violates 
time reversal symmetry \cite{Rashba03}.

\begin{figure}[t]
\onefigure[height=6cm,clip]{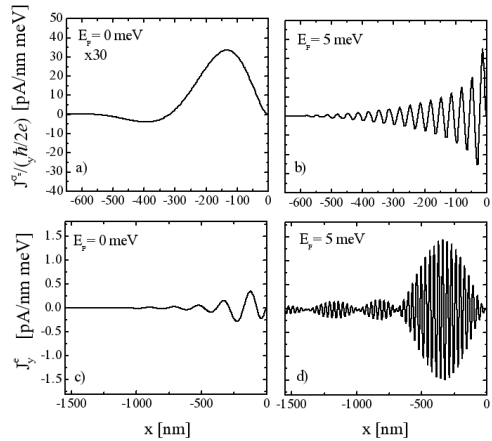} 
\caption{Total spin (a,b) and charge (c,d) current density in the vicinity of
the edge for different values of $E_{\mathrm{F}}$. Notice the
currents are localized near the edge. Here $g\mu _{B}B_{x}=0.25$meV and $\alpha =5$meVnm.}
\label{charge current}
\end{figure}

The counterpart of the current induced magnetization at the sample edge, is
the generation of edge currents by a magnetic field. To avoid the
diamagnetic coupling and Landau quantization, we now analyze the response of
the system to an external in-plane magnetic field. In the continuous model, the
propagating modes are still given by Eq. (\ref{band}) but
the spinor's angle is such that $e^{\ci\phi }=z/|z|$ with $%
z=\alpha (k_{y}-\ci k_{x}) + \frac{1}{2}g\mu _{B}(-B_{x}+\ci%
B_{y})$ and $g$ the g-factor. The presence of the magnetic field modifies
the Fermi surface which is now given by $E_{\mathrm{F}}=\hbar
^{2}k^{2}/2m^{*}\pm |z|$. Notice it is not longer a circle. It is then clear
that the magnetic field introduces an asymmetry between $k_{y}$ and $-k_{y}$
when applied in the $x$-direction. In bulk, the charge
current induced by the field is zero due to a cancellation between the
contributions of the two bands. However, at the edge the evanescent modes
lead to the appearance of a non-zero {\it charge current density}, 
\begin{equation}
J_{y}^{e}(x,\varepsilon )=e\sum_{k_y,\nu}\,\langle v_{y}\rangle
_{k_y,\nu}\,\delta (\varepsilon -\varepsilon_{k_y,\nu})
\label{definition charge 
current}
\end{equation}
The total current density, $J_{y}^{e}(x)$, is obtained by 
integration of this quantity up to $\Ef$. Figs. 3c and 3d shows $J_{y}^{e}(x)$ for $E_{\mathrm{F}}=0$ and $5$meV. As in
the case of the spin current, the charge current is localized at the surface
and presents modulations with two characteristic wavelengths, $\lambda _{F}/2
= 2\pi/(k_{F,+}+ k_{{F},-})$ and $\lambda _{\alpha }= 2\pi/(k_{F,-}- k_{{F},+})$. 

\begin{figure}[t]
\onefigure[height=8cm,clip]{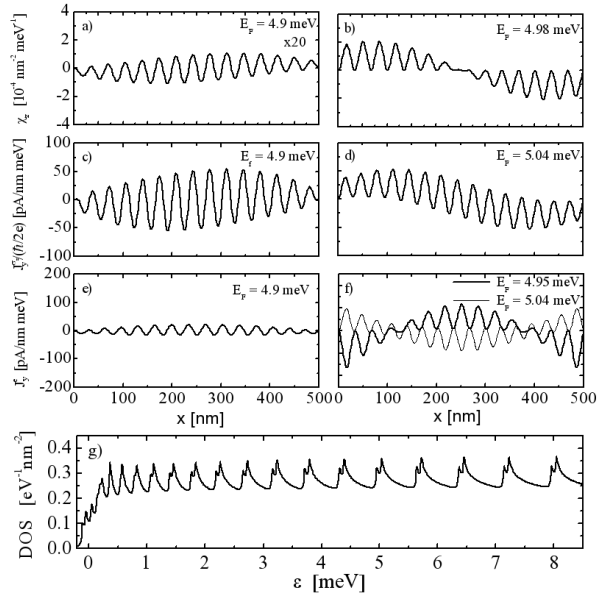} 
\caption{Current-induced spin accumulation (a, b) in the absence of a magnetic field and charge (c, d) and spin (e,f) current density in
the presence of an in-plane magnetic field for a quantum wire.
The panels on the left (a, c and d) correspond to the case of a non-resonant energy, 
$\varepsilon =4.9$meV%
, while the panels on the right (b, d and e) correspond to resonant cases, $\varepsilon =4.98$meV,
$\varepsilon =4.95$meV and $5.04$meV respectively. We used: $g\mu _{B}B_{x}=0.25$meV, $L=500$nm and $\alpha =5$meVnm.
Panel (g) shows the density of states in the presence of the magnetic field. Notice the singularities are split. 
}
\label{finite strip}
\end{figure}

So far we have presented results for wide (large $L_{x}$) systems. We have
shown that out of equilibrium, when a ballistic transport current flows in
the system, 
there is spin accumulation at the
edges. 
The spin and charge current densities discussed above are properties of the ground
state and depend on a delicate balance of the contribution of states with and without evanescent waves. At a given
energy, the number of states of each type can be changed by engineering the 
sample geometry and thus altering the partial cancellation. In particular
this can be done in a narrow (small $L_{x}$)\ quantum wire. The lateral
confinement of the 2DEG in the $x$-direction leads to the quantization of
the transverse modes and the density of states presents quasi-one
dimensional van Hove singularities (see Fig. \ref{finite strip}g). When $E_{\mathrm{F}}$ is tuned to be close to
a singularity, there are many states with $k_{y} \sim 0$---thus with no
evanescent waves---that define the behavior of the system.

The results for the current-induced spin accumulation, obtained using the tight-binding approach, are shown in Figs. \ref{finite
strip}a and \ref{finite strip}b in the absence of a magnetic field. 
Since the 
wave\-func\-tions of the states with $k_y=0$ have a simple form $%
\Psi _{\pm }(x)=L_{x}^{-\frac{1}{2}}\exp (\mp \ci\pi x/\lambda
_{\alpha })\sin (n\pi x/L_{x})\left( 
{\pm \,\ci \atop 1}%
\right) $, where $n$ is the channel
index, one can readily show that, for $k_y \gtrsim 0$, 
$\langle \sigma_z\rangle\propto\sin ^{2}\left(n\pi x/L_{x}\right)$ $ \sin \left(2 \pi(x-L_{x}/2)
/\lambda _{\alpha })\right)$ in excellent agreement with the numerical result of Fig. 4b ($n=15$ for $\Ef\approx5$meV).
In this case, the sign of the magnetization at the edges also depends on the ratio $L_{x}/\lambda_{\alpha}$.
For the parameters used in Fig. 4b the spin density is positive in the interval $[0,L_x/2]$.
Figs. \ref{finite strip}c and \ref{finite strip}d shows the spin current profile.
The resonant case presents well defined signs: there is a net spin current flowing in one
direction on the right and in the opposite on the left. 

The charge current density in the presence of an in-plane field is shown in Figs. \ref{finite
strip}e and \ref{finite strip}f.
For $|B_{x}| \ll  2\alpha k_{\mathrm{F}}/g\mu _{B}$ a similar calculation as above gives  
$\langle v_{y}\rangle_{\pm} \propto\pm
\sin ^{2}\left(n\pi x/L_{x}\right)$ $ \cos \left( 2\pi(x-L_{x}/2)
/\lambda _{\alpha })\right)$.  
Though this expression does not fit the numerical result exactly (Fig \ref
{finite strip}f), it contains the correct spatial modulations. In
particular, it shows that the charge current density
changes its sign in a length scale given by $\lambda _{\alpha }/2$. It is also straightforward
to check that the van Hove singularity is split by the magnetic field---the energy splitting being 
 $\Delta \approx g\mu
_{B}B_{x}|\sin (\pi L_{x}/\lambda _{\alpha })|\lambda _{\alpha }/L_{x}\pi$
for $n \gg  L_{x}/\lambda _{\alpha }$ (see Fig. 4g). 

In summary, we showed that the presence of a potential barrier in systems
with spin-orbit coupling leads to the appearance of 
current-induced magnetization at the sample edges and of equilibrium spin and charge current densities.
These effects are important for a correct interpretation of numerical results in finite systems, where 
geometry plays a crucial role. On the experimental side, low density ballistic 2DEGs or quantum wires are required in order to observe the spin accumulation described here. It is worth mentioning, that these conditions are not met by the recent experiment of Ref. \cite{Sih}, where disorder is very large.
\acknowledgments
We appreciate discussions with B. Alascio, M. J. Sanchez and D. Ullmo and useful comments by E. Rashba. 
This work was partially supported by ANPCyT Grants No 13829 and 13476 and Fundaci\'on Antorchas, Grant 14169/21. 
GU acknowledge support from CONICET.


\begin{thebibliography}{31}
\bibitem{note1}
Note that $\langle \ldots\rangle_{\bm{k},s}$ has to be calculated with the orthogonalized states, see below.
\end{thebibliography}
\end{document}